\documentclass[reprint,amsmath,amssymb,aps,superscriptaddress,floatfix]{revtex4-1}

\usepackage{graphicx}
\usepackage{dcolumn}
\usepackage{bm}
\usepackage{hyperref}
\usepackage{siunitx}
\usepackage{color}
\usepackage{xspace}

\DeclareSIUnit\kT{$k_B T$}
\DeclareSIUnit\dyne{dyne}
\DeclareSIUnit\molar{M}
\newcommand{\kT}{{k_B T}}

\begin{document}

\title{ {\it Volvox barberi} flocks, forming near-optimal, 
two-dimensional, \\ polydisperse lattice packings
}


\author{Ravi Nicholas Balasubramanian}
\affiliation{Harriton High School, 600 North Ithan Avenue, Bryn Mawr, PA 19010, USA
}

\begin{abstract}
{\it Volvox barberi} is a multicellular green alga forming spherical colonies of 10000-50000 differentiated somatic and germ cells.   Here, I show that these colonies actively self-organize over minutes into ``flocks'' that can contain more than 100 colonies moving and rotating collectively for hours.  The colonies in flocks form two-dimensional, irregular, ``active crystals'', with lattice angles and colony diameters both following log-normal distributions.  Comparison with a dynamical simulation of soft spheres with diameters matched to the {\it Volvox} samples, and a weak long-range attractive force, show that the {\it Volvox} flocks achieve optimal random close-packing.   A dye tracer in the {\it Volvox} medium revealed large hydrodynamic vortices generated by colony and flock rotations, providing a likely source of the forces leading to flocking and optimal packing.

\end{abstract}

\maketitle

\section*{Introduction}
The remarkable multicellular green alga {\it Volvox barberi} \cite{kirk2005volvox} forms spherical colonies of 10,000 to 50,000 cells embedded in a glycol-protein based extra cellular matrix (ECM) and connected by cytoplasmic bridges that may have a role in the transport of nutrients and information \cite{hoops2006cytoplasmic}.  Colonies have a surface population of somatic cells that specialize in flagellar beating, and a smaller population of germ cells that undergo cell divisions to form daughter colonies.  Each {\it V. barberi} somatic cell bears two apical flagella projecting outside the colony boundary, and a chloroplast to enable photosynthesis.  The synchronized beating of the flagella of the somatic cells moves the colony forward in a spiral motion -- the colony advances while rotating around a fixed, body-centered axis  \cite{kirk2005volvox}.  Anton van Leeuwenhoek first described {\it Volvox} in 1700, and Linnaeus named them for their hypnotic rolling motion in 1758.  {\it Volvox} is remarkable for the collective behavior and coordination shown by the large number of individual cells within each colony, and is a model organism for studying the transition from uni-cellularity to multi-cellularity with differentiated cell types \cite{kirk2005volvox, herron2016origins, matt2016volvox}.   Strikingly, the coordination of the many cells in a {\it Volvox} colony may be based on hydrodynamic interactions only \cite{brumley2014flagellar,brumley2015metachronal} without chemical signaling between cells. \\

Collective behaviors are mechanisms used by groups of living organisms to support critical functions including self-defense, mating, and predation.  These behaviors occur when a whole group reacts together by means of intercommunication or signaling (see examples in \cite{camazine2003self}).   Examples of collective behavior, at the organismal level, include migration of populations and congregation of individuals for protection, as seen in flocking by birds and schooling by fish (see, e.g., \cite{camazine2003self}).  Microorganisms can also show collective behavior (see, e.g., \cite{drescher2009dancing,petroff2015fast} and references therein) but their interactions are often dominated by viscous forces (e.g., fluid drag) unlike larger organisms which are dominated by inertial forces.  Here, I show that {\it V. barberi} colonies, which are themselves composed of many individual cells acting together, show collective behavior at a higher level of organization.  Entire colonies can dynamically gather into large populations that  move together, even while individual colonies can continue to rotate separately.  When such {\it V. barberi} flocks form, the colonies aggregate into geometric lattices. \\

In mathematics, objects are considered Òoptimally packedÓ when they gather so that as many  fit into a given space as possible.  Optimal packing varies greatly with the shape and size of objects, and with the number of dimensions.  For example, optimal packing of equal-sized squares in a two-dimensional space places them next to each other column by column, row by row. For circles of the same size, optimal packing forms an equilateral triangular lattice, i.e., lines connecting the centers of touching circles form equilateral triangles (see \cite{conway2013sphere}).  In three-dimensional space, equal-sized cubes can stack tightly into larger blocks while uniform spheres pack into a face-centered cubic lattice \cite{hales2005proof}. In other cases, e.g. if circles (in 2d) or spheres (in 3d) are polydisperse, that is, if they have varying sizes, the best packing is not known, but numerical simulations have shown that random close-packed spheres achieve a packing that is nearly optimal \cite{torquato2010jammed}.  I show here that {\it V. barberi} flocks have a log-normal distribution of colony radii, and form lattices which are nearly identical structurally to random close-packings in molecular dynamics simulations of weakly-attracting spheres with the same radius distribution.  The characteristic feature of both the {\it Volvox} and simulation lattices is a log-normal distribution of lattice vertex angles.  This shows that {\it Volvox barberi} gather into flocks that are near-optimally packed for their polydisperse radius distribution.

\vfil

\section*{Results}

\begin{figure}
	\includegraphics[width= \linewidth]{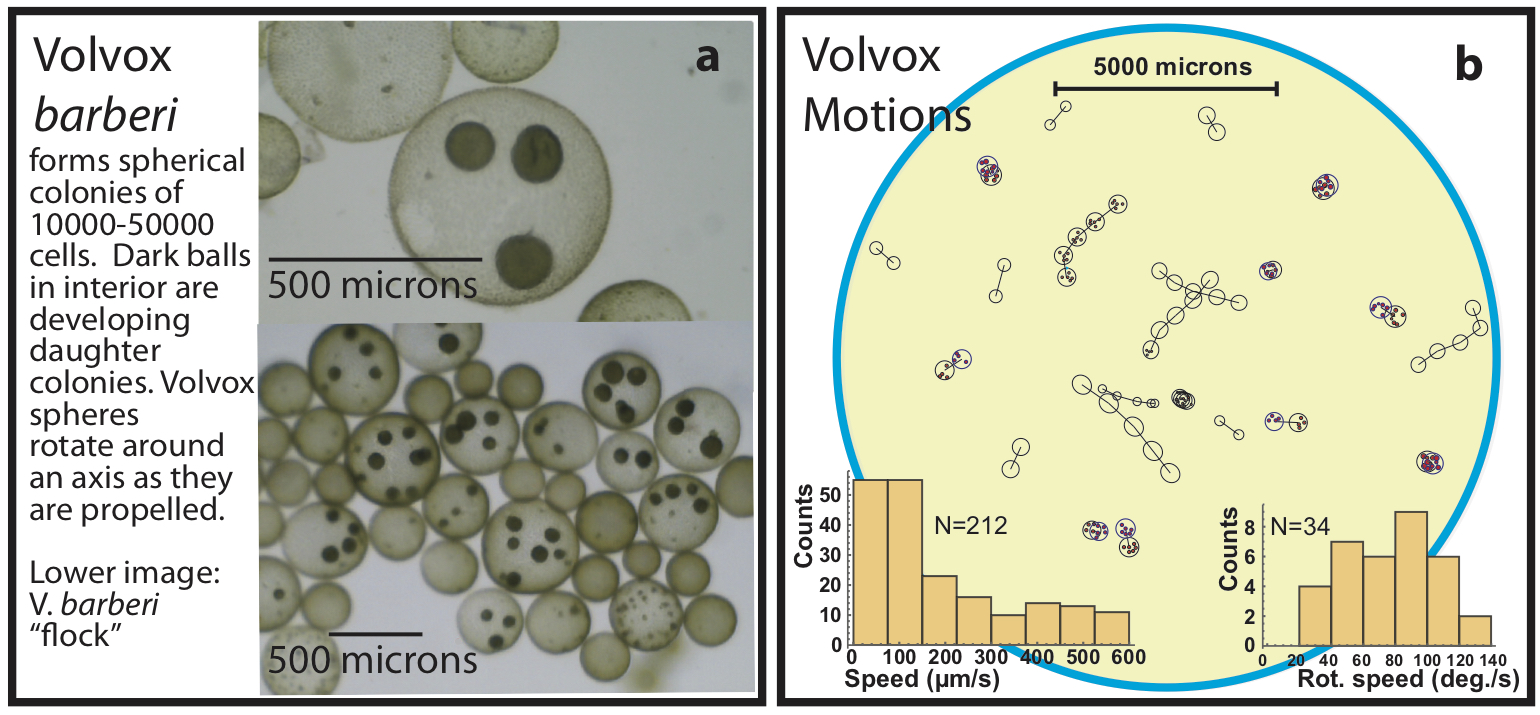}
	\caption{
	\label{fig:motions}
	 ({\bf a}) Images of {\it Volvox barberi}.  ({\bf b}) Schematic composite of {\it Volvox} motions made from time-lapse photographs.  Open circles/ovals are individual {\it Volvox} colonies, solid red circles are unhatched daughter colonies within a mother colony, lines between centers indicate linear trajectories.  Photographs were taken once per second.  Rotational motion in the liquid surface plane was measured by tracking daughter colonies or elliptical shapes.  {\bf Bottom left inset:} Histogram of linear speeds.  {\bf Bottom right inset:} Histogram of rotational speeds.	
	}
\end{figure}

\subsection*{ {\bf {\it Volvox barberi} colonies gather into large flocks }}
\vspace{-0.1in} 
Samples of {\it V. barberi} were placed in culture wells and imaged every second using a high-resolution camera in time-lapse series (see Methods). The motile somatic cells of each colony were organized in a spherical external matrix ranging in diameter from $\sim$$150 \, \mu{\rm m}$ to $\sim$$500 \, \mu{\rm m}$ (Fig.~\ref{fig:motions}a). These colonies moved very quickly, at speeds of up  to $\sim$$600 \, \mu{\rm m/s}$ (Fig.~\ref{fig:motions}b).  This meant that flagellar beating by the somatic cells could move {\it Volvox} colonies by more than a body length each second, the equivalent of 300 km/h for a Boeing 747 jet plane.  While swimming, individual colonies rotated along a body-centered axis.   Rotational speeds were also observed to be very high, as much as $\sim$$120$ degrees/s  (Fig.~\ref{fig:motions}b).  Many colonies moved and rotated individually, but sometimes gathered into remarkable congregations consisting of more than a hundred colonies arranged in stable geometric lattices moving and rotating collectively within the medium (Fig. 1a, bottom).  Individual colonies often continued to rotate separately within these lattices indicating that colonies remained independent, and were not simply stuck or mechanically attached.  This new form of collective organization can be termed a ``flock'', recalling collective behaviors in animals like birds and fish.  Further experiments were carried out to characterize the aggregation dynamics and structure of this new form of micro-organismal self-organization.

\begin{figure}
	\includegraphics[width= \linewidth]{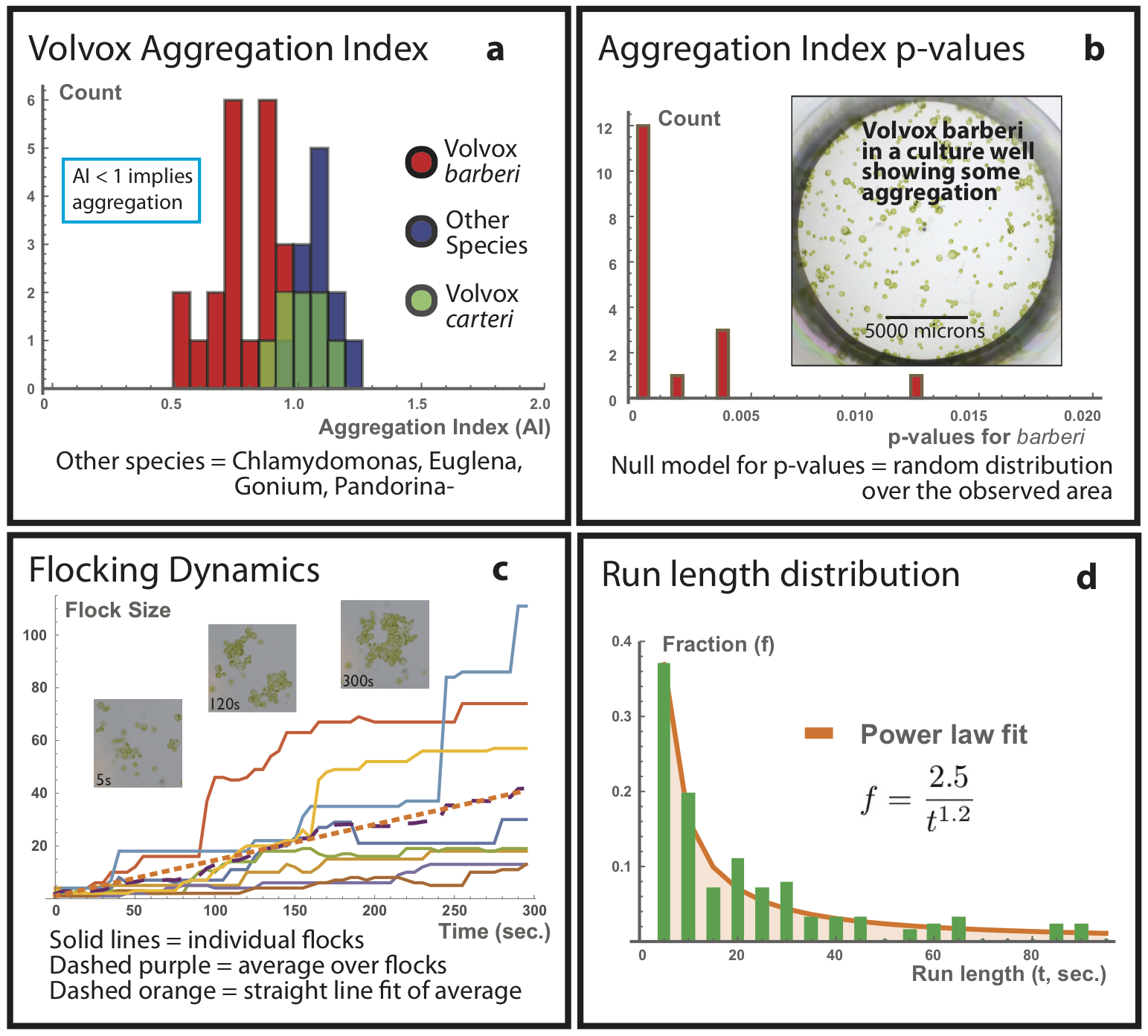}
	\caption{
	\label{fig:aggregation}
	({\bf a}) Aggregation Index (AI, see text for definition) for {\it V. barberi}, {\it V. carteri}, and other species ({\it Chlamydomonas}, {\it Euglena}, {\it Gonium}, and {\it Pandorina}-).  ${\rm AI} < 1$ implies aggregation.  ({\bf b}) p-values for the AI show that {\it V. barberi} is very likely to be aggregating.  ({\bf c}) Flocking {\it V. barberi} groups fluctuate in size and occasionally break apart or merge together.  On average flocks grow at a constant rate of $\sim$0.13 colony/sec (dashed orange line).  Insets show one growing flock at the indicated times.  Measurements of flock size were made in 5 second intervals.  ({\bf d}) Frequencies of run-lengths (duration over which the flock had a constant size) with power law fit.
	}
\end{figure}

\vspace{-0.2in}
\subsection*{Aggregation in {\it V. barberi} cultures}
\vspace{-0.1in}
First,  the Clark-Evans Aggregation Index (AI) \cite{clark1954distance} was applied to test whether {\it V. barberi} colonies in culture wells that did not develop full-fledged flocks were nevertheless attracting each other to some degree.   The Clark-Evans index compares the average distance between nearest neighbors in a sample to the expected nearest-neighbor distance if the organization is random (see Methods).  Thus, ${\rm AI} < 1$ implies aggregation, ${\rm AI} = 1$ implies random organization, and ${\rm AI} > 1$ implies a more evenly spaced, or dispersed, distribution.    Fig.~\ref{fig:aggregation}a compares the measured AI values for samples of {\it Volvox barberi}, and additional species: {\it Volvox carteri}, representatives of two other colonial volvocine genera ({\it Gonium pectorale} and {\it Pandorina morum}), one unicellular volvocine ({\it Chlamydomonas reinhardi}) and one unicellular, non-volvocine protist ({\it Euglena gracilis}). For {\it V. barberi} the AI values were mostly much less than 1, suggesting a strong tendency for the colonies to group together.  By contrast, the {\it V. carteri} AI values clustered around 1, suggesting a random organization, while the other species generally had AI values slightly greater than 1, suggesting a somewhat more even spacing.   Results for the other species confirm that aggregation in {\it V. barberi} is not simply a consequence of the experimental setup.  Clark and Evans also provided a formula to compute a p-value measuring the probability that a random organization would result in a sample with an AI value as measured or more extreme (see Methods).  Almost all the p-values for {\it barberi} were less than $0.001$ (Fig.~\ref{fig:aggregation}b) implying that the results did not arise by chance.

\begin{figure}
	\includegraphics[width= \linewidth]{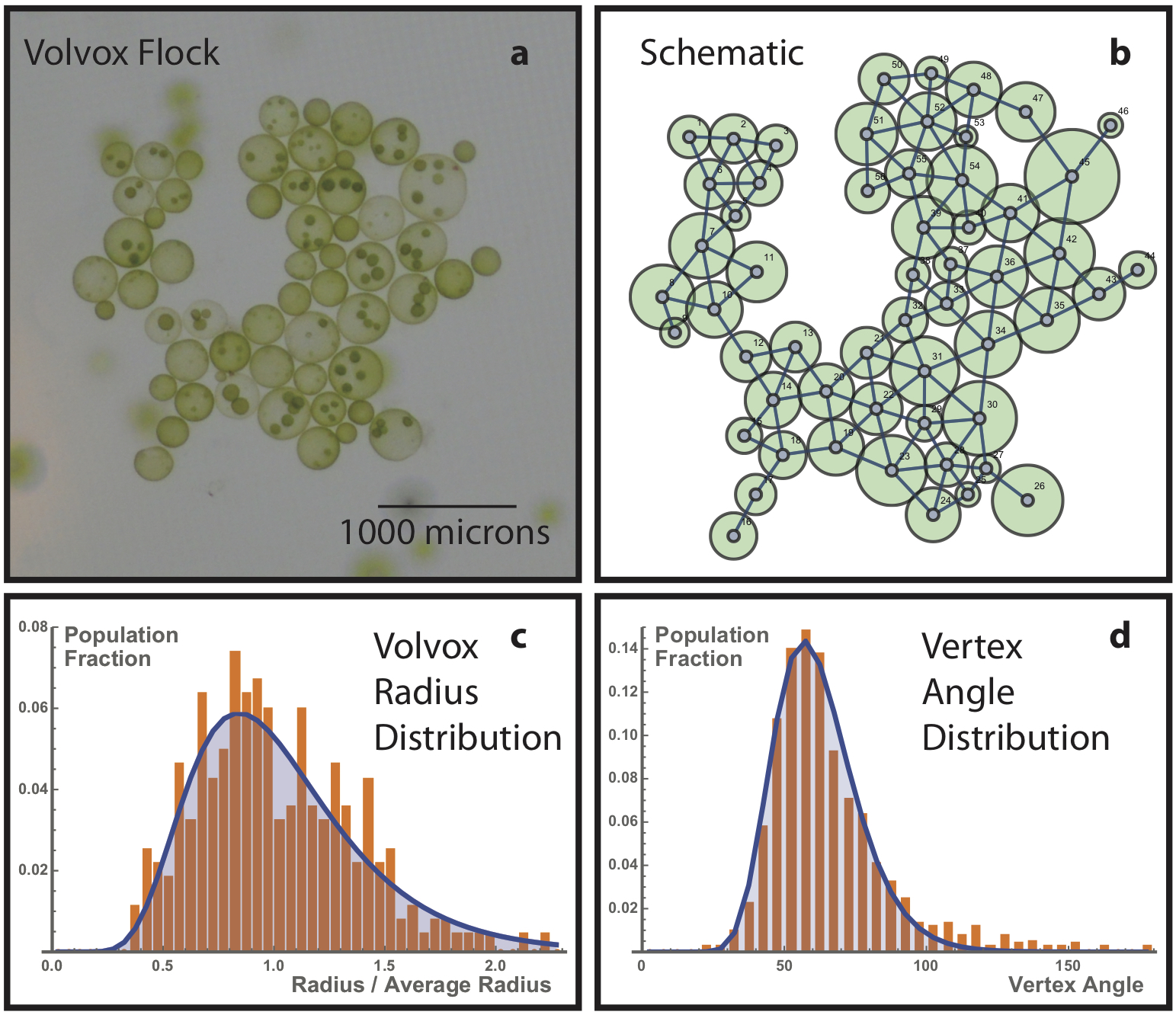}
	\caption{
	\label{fig:geometry}
	({\bf a}) {\it V. barberi} flock where 56 colonies gathered over several minutes and  rotated coherently and rapidly in the culture well.  ({\bf b})  Schematic of flock in panel {\bf a}.  Lines connecting centers of neighboring {\it Volvox} form the flock contact network. ({\bf c}) {\it Volvox} radius distribution (orange bars) fit by a log-normal curve (blue line; log standard deviation $\sigma \approx 0.38$ and log mean $\mu \approx -0.032$; lognormal function defined in Methods).  ({\bf d})  Angle distribution  in the contact network of {\it V. barberi} flocks (orange bars) fit by a log-normal curve (blue line); log standard deviation $\sigma \approx 0.24$ and log mean $\mu \approx 4.10$, corresponding to a mean vertex angle of 64.9 degrees.	
	}
\end{figure}

\subsection*{Dynamics}
\vspace{-0.1in} 
Some samples continued to show strong aggregation over extended periods of time, and eventually gathered into large two-dimensional sheets --  flocks.  The dynamics of the flocking process were measured using time-lapse photography (see Methods).  Over 300 seconds period, flocks changed in shape and size, starting at 2-4 colonies per flock and growing to sizes above 100 (Fig.~\ref{fig:aggregation}c).  The fraction $f$ of run-lengths (seconds of constant flock size) of duration $t$ was well described by a power law as $ f = 2.5/t^{1.2} $ (determined by a least-squares fit, Fig.~\ref{fig:aggregation}d). During periods of growth, individual colonies joined the flock or whole flocks joined together to make a single, bigger, flock.  Occasionally, individual colonies broke from the main flock to make space for another colony, and move to another location along the perimeter of the flock. On average, flocks grew at a rate of $\sim$0.13 {\it Volvox} colonies per second. Once a large flock had formed it could persist for hours before separating again into individual colonies. \\

\begin{figure}
	\includegraphics[width= \linewidth]{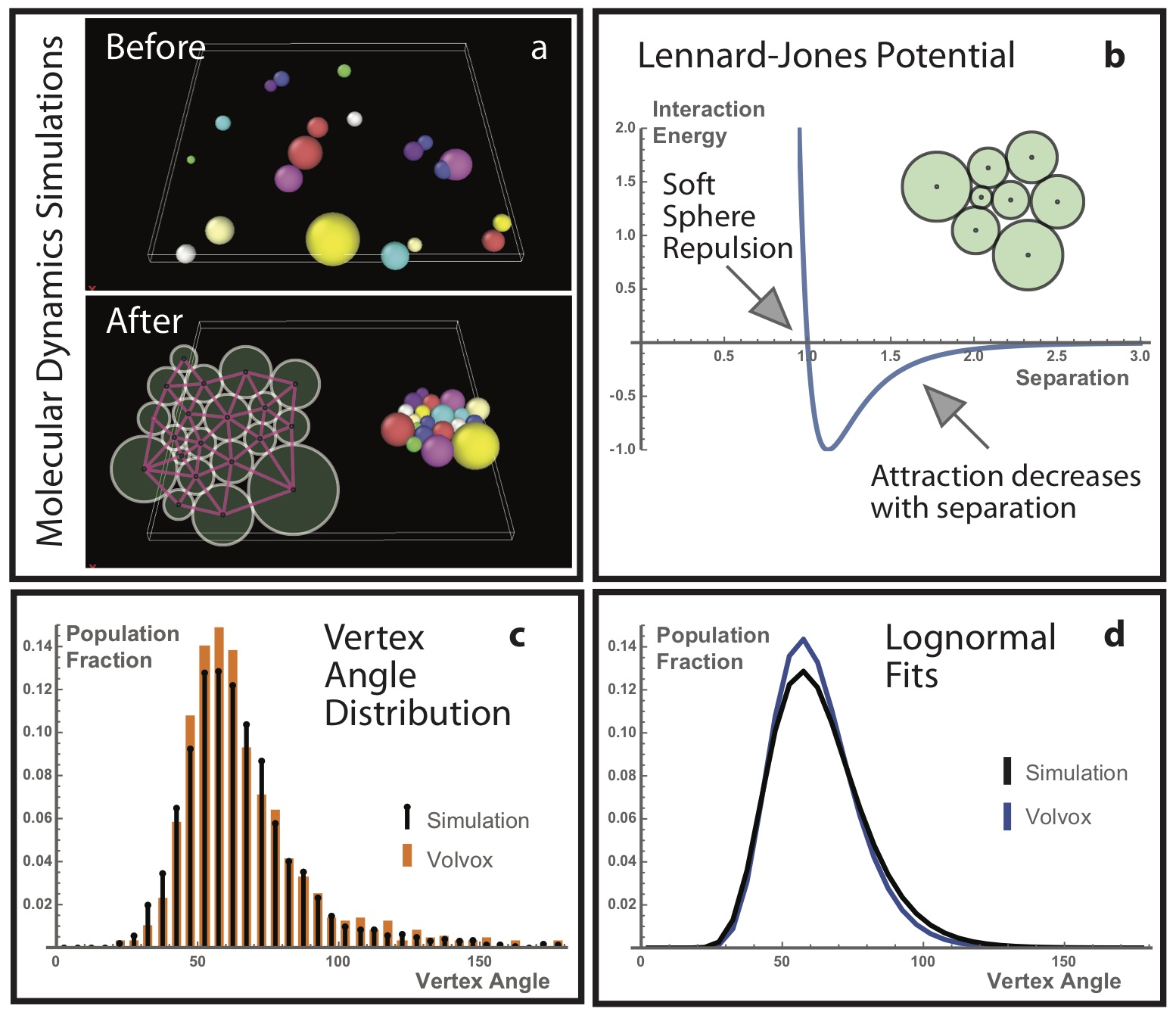}
	\caption{
	\label{fig:optimal}
	({\bf a}) Dynamical simulations of weakly attracting soft spheres moving in two dimensions. Random initial configurations (Before) led eventually to packed lattices (After).  Sphere radii were matched to the radii of colonies in {\it Volvox} flocks.  Pink lines in the schematic show the final contact network.  ({\bf b}) Interaction energy between spheres (here with radii summing to 1) at different distances. Short range repulsion is strong (large positive energies) and long range attraction is weak (small negative energies).  Inset schematic shows a small final configuration.  ({\bf c}) Contact network angle distribution in the simulation  matches the {\it V. barberi} distribution (Fig.~\ref{fig:geometry}d).  ({\bf d}) A log-normal fit of the simulation angle distribution (black line; log standard deviation $\sigma \approx 0.26$ and log mean $\mu \approx 4.11$ , corresponding to a mean vertex angle of 64.8 degrees), matches the {\it Volvox} fit (parameters in Fig.~\ref{fig:geometry}).
	}
\end{figure}

\vspace{-0.2in} 
\subsection*{Geometry of large flocks} 
\vspace{-0.1in} 
 As flocks grew, they formed fixed, almost crystalline, two-dimensional lattices whose structure remained stable despite the overall motion of the flock and the separate rotations of individual colonies in it (Fig.~\ref{fig:geometry}a).  To characterize the geometry of these lattices,  a {\it contact network} was constructed -- a lattice formed by points at the centers of {\it Volvox} colonies with lines connecting touching pairs (Fig.~\ref{fig:geometry}b).  The contact diagram is closely related to the Delaunay tessellation of packed spheres, and simply removes lines in the tessellation that do not run between touching elements. Contact networks are known to be important in determining when a packing of spheres is rigid \cite{torquato2010jammed}.  In the contact networks of {\it Volvox} flocks most lattice cells were triangular, with the occasional appearance of other polygons (such as quadrilaterals and pentagons).  As seen in Fig.~\ref{fig:geometry}b, the {\it Volvox} were polydisperse (i.e. have varying radii) which affects the lattice structure. The distribution of the relative radius (x = radius/average radius) was skewed towards large values and was well-described by a log-normal function (Fig.~\ref{fig:geometry}c; see Methods).  Similarly, the distribution of angles at vertices in the contact network was skewed towards larger angles and was also well-described by a log-normal function (Fig.~\ref{fig:geometry}d).  When viewed with the naked eye, colonies in {\it Volvox} flocks seemed as tightly organized as possible.

\subsection*{Optimal two-dimensional polydisperse sphere packing in {\it Volvox} flocks}
\vspace{-0.1in} 
Optimal packing is the property of objects being as tightly organized into a given space as possible, e.g., packing the most oranges into a crate.   Optimal packing of objects depends on their size and shape distributions and the number of dimensions in which they are assembled.    Although {\it Volvox} are three-dimensional, their flocks formed two-dimensional sheets, so the relevant packing problem is for circles ({\it Volvox} cross-sections).  However, even in two dimensions the optimal packing of circles of two or more sizes is unknown.  Random close-packing achieves an excellent approximation to the optimum  \cite{torquato2010jammed}.

To determine the random close-packed organization of spheres with diverse sizes like {\it Volvox} (Fig.~\ref{fig:geometry}c),  a molecular dynamics simulation was constructed (see Methods).  Each simulation was initialized with randomly placed spheres matched to the size distribution of each flock (Fig. 4a, Before), and allowed the spheres to interact via weak pairwise long-range attractive forces determined by Lennard-Jones potentials (Fig.~\ref{fig:optimal}b; Methods).   At short distances the spheres experienced repulsion.  Pairs attained their lowest energy when they were separated by the sum of their radii.  These Òsoft spheresÓ moved in two dimensions with random initial velocities and were slowly ÒcooledÓ by the simulation software so that their mutual attraction settled them into a packed configuration (Fig.~\ref{fig:optimal}a, After).  The resulting angle distribution for the contact diagrams of the jammed final configurations matched the {\it Volvox} angle distribution (Fig.~{\ref{fig:optimal}c).  This was confirmed by fitting a log-normal curve to the simulation angles -- the fit parameters (Fig.~\ref{fig:optimal}d caption) were almost identical to the parameters of the fit to the {\it Volvox} angles (Fig.~\ref{fig:geometry}d, caption). This suggests that {\it Volvox barberi} flocks achieve random-close packing, which in general gives the closest known approximation tooptimal packing of polydisperse spheres.

\begin{figure}
	\includegraphics[width= \linewidth]{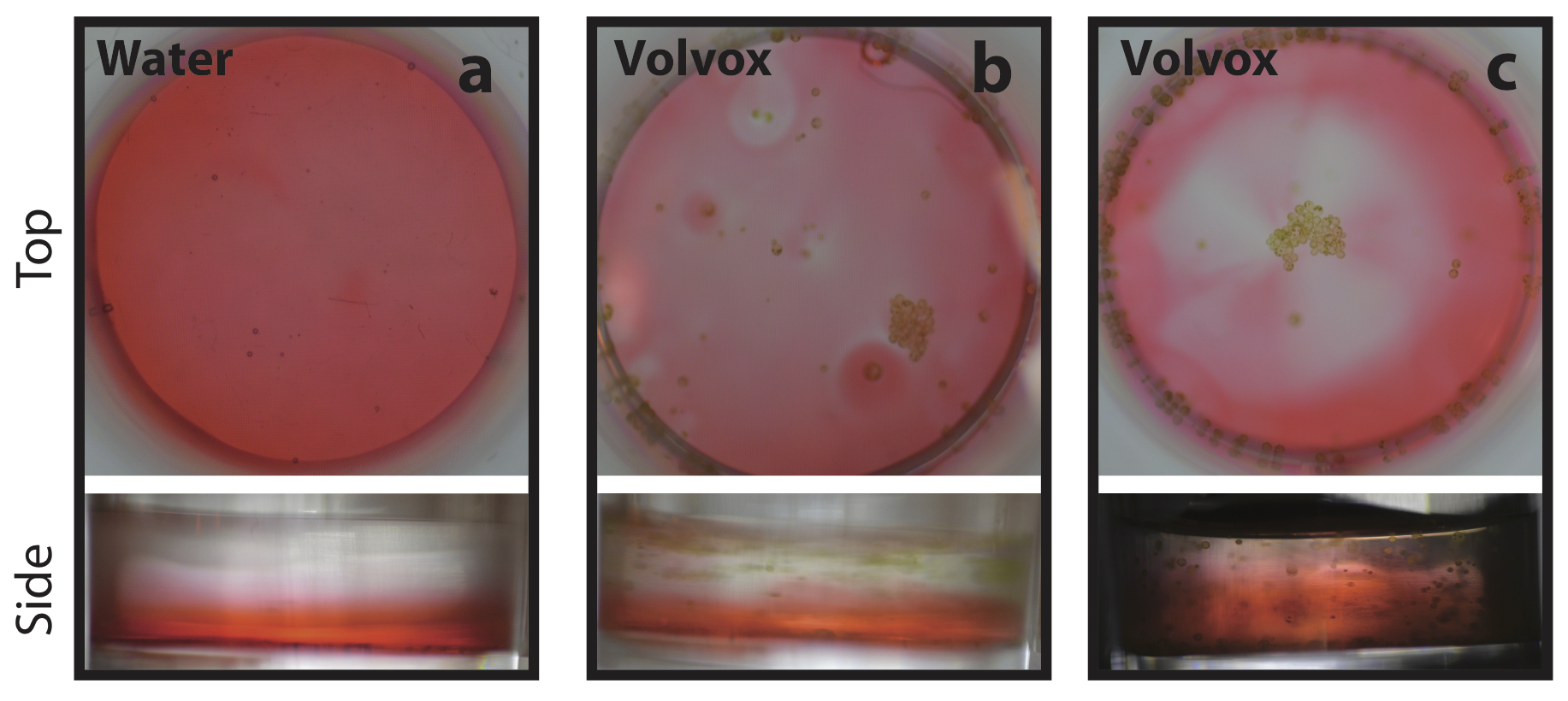}
	\caption{
	\label{fig:vortices}
	({\bf a})  {\bf Top Panel}: Culture well with water and dye tracer after five minutes viewed from above.  Dye is spread out evenly across culture well. {\bf Bottom panel}: Same culture well viewed from the side.  Dye forms a uniform layer at the bottom of the well.  ({\bf b}) {\bf Top panel}: Medium with {\it Volvox} and a dye tracer viewed from above.   Numerous streaks and patches created by {\it Volvox} motions are visible. A rapidly rotating pair of {\it Volvox} near the top forms a clearing in the liquid, while an already formed flock moves slowly at bottom right clearing a channel.  {\bf Bottom panel}: Dye forms a uniform layer at the bottom of the culture well.   ({\bf c}) {\bf Top panel}: {\it Volvox} in medium with a dye tracer viewed from above.  A large, rotating flock formed and created a clearing in the liquid that grew over time, here termed a {\it vortex}.  Diffuse spindles from the edge to the center developed at late times. {\bf Bottom panel}: Same culture well viewed from the side.  Despite the large vortex, the dye is uniformly spread as seen from the side suggesting that the clearing in the upper panel is a cylindrical vortex.
	}
\end{figure}

\vspace{-0.2in} 
\subsection*{{\it Volvox} vortices: a candidate mechanism for flocking}
\vspace{-0.1in} 
What mechanism enables {\it Volvox} to flock?  It is known that another species of {\it Volvox} -- {\it V. carteri} Ð is capable of using fluid forces created by flagellar beating to form Òwaltzing pairsÓ \cite{drescher2009dancing}.  I hypothesized that V. barberi, being one of the fastest species of {\it Volvox} \cite{solari2008volvox}, could exert similar attractive forces  strong enough to pull {\it V. barberi} into well-packed sheets.  To test for these attractive forces, a dye tracer was placed into culture wells containing either {\it Volvox} in medium or clear water (as a control).  The dye was left to dilute over the course of 5 minutes, while the culture wells were imaged every second from above.  After five minutes the final state of the well was recorded from above and from the side (see Fig.~\ref{fig:vortices}). In every control with just water, the dye spread evenly across the culture well when viewed from above, while also forming a uniform layer at the bottom of the well when viewed from the side (Fig.~\ref{fig:vortices}a, upper and lower).  In culture wells with {\it Volvox}, however, movements of the colonies created streaks and patches in the liquid, clearings where the dye tracer was pushed out (Fig.~\ref{fig:vortices}b and c). When large, rotating flocks formed, they created large, almost circular clearings in the liquid that spanned many flock diameters, here termed vortices (Fig.~\ref{fig:vortices}c).  When the {\it Volvox} culture wells were viewed from the side, there was again a uniform band of dye at the base of the well.  This showed that the large vortices associated to flocks were cylindrical.  In some cases, it was possible to see spindles of dye pointing towards the center of the flock. These spindles may be channels where fluid drawn into the flock vortex was pulling dye from the periphery into the center, where it was spun around before being evenly diffused back into the liquid.  The same forces could pull free-swimming {\it Volvox} colonies into the flock.

\section*{Discussion}
This paper showed that {\it Volvox barberi}, a multicellular colonial alga, is capable of a remarkable form of dynamical self-organization, here called flocking, where as many as a hundred {\it Volvox} colonies gather into packed two-dimensional lattices over a period of minutes.  Flocks grow in size as individual colonies attach to them, or as two flocks join.  Once formed, flocks can last for more than an hour before they break apart again into individual colonies.  The {\it Volvox}, which have a lognormal radius distribution, form a contact network composed mostly of triangles, which in turn have a lognormal vertex angle distribution.  A molecular dynamics simulation showed that {\it Volvox} achieve a random close-packing which is nearly optimal for the radius distribution.  

In larger species such as birds it has been shown that weak signals between pairs of organisms can be sufficient to organize the complete flock \cite{bialek2012statistical}.  Here dye-tracing  demonstrated that rotating {\it Volvox barberi} flocks generate vortices that range over many flock diameters, providing a possible mechanism for aggregation of colonies.  Drescher et al. showed that colonies of another species, {\it Volvox carteri}, use fluid flows generated by flagellar beating to form Òwaltzing pairsÓ near the boundaries of culture wells \cite{drescher2009dancing}. Similarly, it could be that {\it Volvox barberi} rotating in the same direction produce vortices whose flows cancel near the midpoint between colonies leaving a space that the {\it Volvox} can easily move into.  Alternatively, counter-rotating {\it Volvox} may produce faster fluid flows in the space between them; by the Bernoulli principle (which also controls airplane wing design) the faster moving fluid would create a lower pressure causing  the slower exterior fluid to push colonies together.  These two alternatives could be distinguished by careful imaging of attracting pairs, and by theoretical models of the hydrodynamics of fluid flows created by {\it Volvox} colonies.

Recently, Petroff et al. reported that a fast-moving, single-celled bacterium, {\it Thiovulum majus}, can form Òactive two-dimensional crystalsÓ \cite{petroff2015fast}.  {\it Thiovulum} are spheres of a stereotyped diameter (8.5 $\mu$m).  Thus their Òactive crystalsÓ are arranged in a regular hexagonal packing (i.e. an equilateral triangular contact network).  Like {\it Volvox}, {\it Thiovulum} rotates as it moves, and the resulting fluid drag is thought to create tornado-like flows that form the active crystal.  {\it V. barberi} colonies are much larger (up to 500 $\mu$m in diameter) and contain tens of thousands of beating flagella; hence they can potentially generate much more powerful fluid flows.  Importantly, {\it Volvox} colonies have varying diameters, perhaps because each colony may be in various stages of its life cycle of producing and dispersing daughter colonies.  Because of this polydispersity it is impossible for {\it Volvox} to arrange into a regular lattice. Instead, experiments here showed that the {\it Volvox} lognormal radius distribution gave rise to a lognormal angle distribution in the contact diagram, which matches the outcome of random close-packing.  Thus {\it Volvox barberi} may provide an experimental model for studying how optimal packing can be dynamically achieved in nature.  For example, the dependence of the optimal lattice on the size distribution could be tested by selecting colonies of specific sizes from a sample and then repeating the experiments in this study.

Why should {\it Volvox barberi} flock?  Flocking may be a passive consequence of fluid flows produced by moving {\it Volvox} when they are placed in the artificial environment of a culture well.  However, this seems unlikely since {\it V. carteri}, which was also included in this study, and is known to produce strong hydrodynamic flows, did not flock.  Additionally, {\it V. barberi} colonies continued to rotate and move independently of their flock, and sometimes broke off, moved and reattached elsewhere where they fit better.  This dynamic re-organization to produce better packed lattices suggests that flocking involves more than the accidental cohesion of {\it Volvox} that happen to be in proximity. Alternatively, flocks may form because of forces arising simply from the tension and meniscus shape of the surface of the liquid in the culture well.  This also seems unlikely since the focusing depth of the camera indicated that flocking tended to occur at levels below the surface.  Furthermore, colonies were always able to easily break off from the main flock without adhering to the surface of the medium.  We can test if meniscus shape has an effect on flocking by placing cultures of {\it V. barberi} in containers with varying shapes, sizes and affinities for the medium in order to vary the geometry of the liquid surface.  In addition, the surface tension could be changed by modifying viscosity of the medium.  If a trend is observed where fewer or more flocks form in containers with a greater surface tension or more strongly curved meniscus geometries, then it is likely that surface tension has an effect.

On the other hand, flocking may have some functional value.  For example, a spinning flock could stir the liquid to create a large mixing zone, which would be beneficial for acquiring soluble nutrients.  There may also be a kind of Òsafety in numbersÓ in the face of predation from rotifers, or other stresses.  This could be tested by measuring if flocking occurs more often, or for more prolonged periods of time, if predators are present in the culture well or under conditions of stress.  Another function of the observed flocking might be collective phototaxis in which multiple light-seeking {\it Volvox} colonies associate together to increase their overall buoyancy or to maintain stability in the face of random fluid motions.  This would enable the flock to rise and position itself near the surface more effectively, so that the population as a whole would have access to as much light as possible in order to photosynthesize.  To test this, we could compare rates of flocking at different light exposure levels, or by changing the angle at which light enters the culture well. It would be fascinating to establish whether the visually beautiful phenomenon of {\it Volvox} flocking serves an adaptive function.

\section*{Methods}

\paragraph*{{\bf Strain maintenance: }}
{\it Volvox barberi}, {\it Euglena gracilis}, {\it Chlamydomonas reinhardi}, {\it Gonium pectorale}, and {\it Pandorina morum (-)} were acquired from Carolina Biological and maintained in Woods Hole growth medium \cite{stein1973handbook}.  {\it V. carteri} (Eve strain, HK10) was provided by Dr. James Umen of the Donald Danforth Plant Science Center, St. Louis, MO.  Samples were maintained at room temperature ($\sim$70 degrees Fahrenheit) in sterile, air-filtered culture flasks, under fluorescent illumination from below with a 12/12 day/night cycle. \\

\paragraph*{{\bf Imaging and time-lapse photography: }}
Prior to imaging, 0.3 ml of a sample was placed in a plastic culture well (ThermoFisher Scientific) measuring 1.5 cm wide.  Illumination was provided by a white LCD screen placed below the culture well.  Imaging was carried out from above with a Nikon D3200 camera equipped with a macro-lens and stabilized on a tripod.  Time-lapse sequences were set to take an image every one or five seconds for 300 seconds.  To measure flocking dynamics, the largest developing flock was followed.  Still images of wells without evident flocks were used for the Aggregation Index analysis. \\

\paragraph*{{\bf Clark-Evans Aggregation Index: }}
To measure aggregation, photographs were analyzed using GraphClick, a freely available image analysis program.  Each image was calibrated to convert pixel distances into microns by comparison with the known diameter of the culture well (1.5 cm).  Points were marked on the center and edge of each visible colony.  Colony radii were measured as Euclidean distances between center and edge points on each colony.  Nearest neighbor distances were measured between centers of closest colony pairs. The Clark-Evans Aggregation Index \cite{clark1954distance} was calculated as:
\begin{equation}
r_a = { \sum_{i=1}^n r_i \over  n } ~~;~~ r_e = {1 \over 2 \sqrt{\rho}} ~~;~~ {\rm AI} = {r_a \over r_e}
\end{equation}
where $r_a$ is the average distance to the nearest neighbor, $r_e$ is the expected distance to nearest neighbor if the organization were random, and ${\rm AI}$  is the Aggregation Index.  These quantities are computed from the number of {\it Volvox} ($n$), the distance from the colony indexed by $i$ to its nearest neighbor ($r_i$), and the two-dimensional population number density of the {\it Volvox} ($\rho$).  Clark and Evans (1954) also provided a formula for computing a p-value measuring the probability that random organization would give rise to a sample with an ${\rm AI}$ which is more extreme that a given value.  To compute the p-value I first calculated a z-score  \cite{clark1954distance}:
\begin{equation}
z = { r_a - r_e   \over   0.26}   \sqrt{n\rho}
\end{equation}
A standard statistical package in the Mathematica (Wolfram) computing environment converted the z-score into a probability. \\

\paragraph*{{\bf Flock Geometry: }}
I characterized the geometry of flocks in terms of the {\it Volvox} colony radius distribution, and the vertex angles of the {\it Volvox} contact network.  Radii were measured as described in the Aggregation Index methods above. Custom software (available upon request) produced a contact network consisting of lines connecting the centers of touching {\it Volvox}.  To determine touching pairs, the software checked if the distance between colonies was less than or equal to the sum of their radii. Occasionally, small errors in the marked locations of colony centers (see above) led to extra or omitted lines in the contact diagram; these were corrected through manual curation. Vertex angles were computed by using the cosine rule:
\begin{equation}
c^2 = a^2 + b^2 - 2 a b \cos(\Delta)
\end{equation}
where $a$ and $b$ are the lengths of two adjacent lines ($A$ and $B$) of the contact diagram forming a vertex with an interior angle $\Delta$.  If $A$ and $B$ belong to a triangle in the contact network, then $c$ is the length of the opposite side from angle $\Delta$.  If $A$ and $B$ belong to a higher polygon, the polygon is first triangulated, and $c$ is the length of the line that would be opposite $\Delta$ in an appropriate triangulation. \\

\paragraph*{{\bf Lognormal distributions: }}
The radius and angle distributions were fit with log normal distributions
\begin{equation}
\Pr(x) = {1 \over \sqrt{2\pi} x \sigma}  e^{-  {  ( \ln(x) - \mu )^2  \over 2 \sigma^2}}
\end{equation}
where $x$ is the variable being measured, and $\mu$ and $\sigma$ are respectively the mean and standard deviation of the natural log of $x$ computed in this distribution.  The mean of $x$ is $e^{(\mu + \sigma^2/2)}$.  The lognormal distribution is skewed towards large values of $x$ like the radius and contact network vertex angle distribution.  Fits were carried out by minimizing the squared difference between measured data and this function. \\

\paragraph*{{\bf Simulations: }}
Computational modeling was carried out in the freely available LAMMPS molecular dynamics software package.  Each simulation involved many circular particles in two dimensions, with radii matched to the {\it Volvox} in a given flock.  The particles attracted each other through Lennard-Jones pairwise potentials (Fig.\ref{fig:optimal}b), described by the equation
\begin{equation}
V(r) = 4\epsilon  \left( {\beta^{12} \over r^{12}} - {\beta^6 \over r^6}      \right)
\end{equation}
Here $\epsilon$ sets the arbitrary units of potential energy;  $\epsilon$ was set to $1$. The potential vanishes when $r=\beta$, and is minimized when $r = 2^{1/6}   \beta \approx 1.12 \,  \beta$.  I carried out the studies in two dimensions, rather than three, because inspection showed that the {\it Volvox} were flocking in a fixed plane.  The Lennard-Jones potential is repulsive at short distances (potential decreasing with distance), attractive above a critical distance (potential increasing with distance), and vanishes at large separations. This potential was originally designed to describe interactions between neutral atoms \cite{jones1924determination}, but was used here as a simple model of weak long-range attraction with short-range repulsion.  The LAMMPS input scripts permit choosing separate parameters for the interaction potentials of each pair of particles.  These parameters were chosen to ensure that an isolated pair would come to equilibrium at late times at a separation equal to the sum of their radii.  This was achieved by fixing  in the equation above so that the pairwise potentials were minimized at the sum of the corresponding radii.  Simulations were initialized with particles at random positions, random initial velocities and a finite temperature.  The simulation cooled the particles slowly so that they could rearrange themselves while attracting each other as they approached equilibrium.  These dynamics are not intended to precisely model the forces exerted by {\it Volvox} on each other, but rather to provide a generic setting where polydisperse, two-dimensional spheres achieve close packing by attracting each other weakly over long distances (here through the Lennard-Jones potential) while rearranging themselves (here through fluctuations induced by finite temperature). \\

\paragraph*{{\bf Dye tracer studies of {\it Volvox} vortices: }}
Red food coloring (FD\&C Reds 3 and 40 suspended in water, propylene glycol, and propylparaben; McCormick \& Co.) was diluted in a proportion of about one drop of coloring per ml of water.  Three drops of diluted dye mixture were placed, using a pipette, at the edge of a culture well that contained either {\it Volvox} in medium or a similar volume of clear water as a control.  Time-lapse photography then imaged the culture well.  The camera was typically positioned above the sample to study how {\it Volvox} motions affected the transverse spread of the dye.  Images were also taken from the side to study how the dye spread vertically.

\begin{acknowledgments}
I am grateful to Richard McCourt of the Academy of Natural Sciences of Drexel University for introducing me to {\it Volvox} and for mentoring this project.  Tristan Sharp, of the University of Pennsylvania, taught me about the LAMMPS molecular dynamics software.  James Umen and Sa Geng of the Donald Danforth Plant Science Center sent me samples of {\it Volvox carteri} and their growth medium.   I also thank all the participants of the 2017 Volvox conference, and especially Matthew Herron and Stephanie Hoehn, for insightful remarks.   Presentations of this work at Volvox 2017 and the 2017 Annual Meeting of the Phycological Society of America were supported by Hoshaw Travel Award of the Phycological Society of America (PSA) and by the Donald Danforth Plant Science Center in St. Louis, MO.
\end{acknowledgments}


%

\end{document}